# Tiled Array of Pixelated CZT Imaging Detectors for *ProtoEXIST2* and *MIRAX*-HXI

Jaesub Hong, Branden Allen, *Member, IEEE*, Jonathan Grindlay, Barbara Rodrigues, Jon Robert Ellis, Robert Baker, Scott Barthelmy, Peter Mao, Hiromasa Miyasaka, Jeff Apple

*Abstract*–We have assembled a tiled array (220 cm$^2$) of fine pixel (0.6 mm) imaging CZT detectors for a balloon borne wide-field hard X-ray telescope, *ProtoEXIST2*. *ProtoEXIST2* is a prototype experiment for a next generation hard X-ray imager *MIRAX*-HXI on board *Lattes*, a spacecraft from the Agencia Espacial Brasilieira. *MIRAX* will survey the 5 to 200 keV sky of Galactic bulge, adjoining southern Galactic plane and the extragalactic sky with 6′ angular resolution. This survey will open a vast discovery space in timing studies of accretion neutron stars and black holes. The *ProtoEXIST2* CZT detector plane consists of 64 of 5 mm thick 2 cm × 2 cm CZT crystals tiled with a minimal gap. *MIRAX* will consist of 4 such detector planes, each of which will be imaged with its own coded-aperture mask. We present the packaging architecture and assembly procedure of the *ProtoEXIST2* detector. On 2012, Oct 10, we conducted a successful high altitude balloon experiment of the *ProtoEXIST1* and *2* telescopes, which demonstrates their technology readiness for space application. During the flight both telescopes performed as well as on the ground. We report the results of ground calibration and the initial results for the detector performance in the balloon flight.

## I. Introduction

A main challenge for future hard X-ray (~5 to 200 keV) survey telescopes comes from a need for large area hard X-ray sensors with fine spatial resolution. CZT detectors are promising candidates. The Hard X-ray Imager (HXI) on Monitor e Imageador de RAios-X (*MIRAX*), a next generation wide-field hard X-ray survey telescope, will make use of recent advances in CZT detector technology. *MIRAX*, scheduled for launch in 2017, is the first Brazil space astrophysics experiment on the *Lattes* satellite from the Instituto Nacional de Pesquisas Espaciais (INPE) (Fig. 1) [1].

The main science objectives of *MIRAX*-HXI are 1) systematic survey of the full Galactic bulge and southern Galactic plane to understand the distributions and nature of accreting compact sources including transients and obscured isolated black holes in giant molecular clouds, 2) exploration of the rich time-domain astrophysics including jets and bursts from accretion compact objects and active galactic nuclei, and 3) joint investigation of short gamma-ray bursts in hard X-rays with gravity waves (e.g. Advanced LIGO) [2].

*MIRAX*-HXI will consist of identical 2×2 coded-aperture telescopes with 256 cm$^2$ CZT detector plane in each telescope, providing a combined Field of View (FoV) of 50⁰ × 50⁰ fully coded (a section of the sky accessible by every part of the active detector area through the mask) and 62⁰ × 62⁰ at 50% coding (a section of the sky accessible by only 50% of the detector area through the mask). It will scan a half of the sky with 6′ angular resolution in the 5 to 200 keV band. The HXI will point at 25⁰ south of the zenith to efficiently survey the Galactic bulge and adjourning southern Galactic plane from a low inclination low Earth orbit (650 km, 15⁰ inclination). The wide FoV will also cover a large portion of the extragalactic sky. The modular design with a relatively small FoV of each telescope allows relatively low background, compared to a single telescope design with a similar overall FoV. With the low threshold (5 keV) and low background, *MIRAX*-HXI can achieve the 2.5 to 4 times better sensitivity than its predecessors *INTEGRAL*/IBIS [3] and *Swift*/ BAT [4], which had low energy thresholds of 17 and 15 keV respectively.

In order to localize new hard X-ray sources within ~1′ precision without relying on another softer X-ray telescope, sub-mm spatial resolution is required for the HXI detectors at ~70 cm separation between the mask and detector plane. At the same time, in order to achieve sub-mCrab sensitivity over the scanned half of the sky in a 1 year survey, the active area of the detector has to be ~1000 cm$^2$ or larger. The low threshold (5 keV), which allows a nice overlap with the energy coverage of many soft X-ray telescopes, is needed for the full energy coverage of both thermal and non-thermal processes. This means that energy resolution of the detectors has to be 2 keV (FWHM) or better. Advanced CZT imaging detectors that we have been developing under our balloon-borne program *ProtoEXIST1* (*P1*) & *2* (*P2*) [5,6,7] will meet all the above requirements for *MIRAX*-HXI. In addition to lab experiments to verify the performance of these detectors, we have also conducted a series of high-altitude balloon flights and demonstrated their technology readiness for space application successfully.

This paper is organized as follows. We introduce our CZT detector development program *P1* & *P2* and their telescope design in section II. In section III, we present the design architecture and assembly procedure of the *P2* CZT imager. In section IV, we show pre-flight boresight calibration results. In section V, we describe a successful high-altitude balloon flight experiment for both the *P1* & *P2* telescopes on 2012 Oct, 10. Finally, we show the inital results of the *P2* detectors from the

Manuscript received November 16, 2012. This work was supported by NASA grants NNX09AD96G and NNX11AF35G.

JaeSub Hong is with the Harvard-Smithsonian Center for Astrophysics, Cambridge, MA 02138 USA (telephone: 617-496-7512, e-mail: jaesub @head.cfa.harvard.edu). B. Allen, J. Grindlay and J. Ellis are also with the Harvard-Smithsonian Center for Astrophysics (CfA), Cambridge, MA 02138 USA. B. Rodrigues is with the Instituto Nacional de Pesquisas Espaciais (INPE), Brazil.
R. Baker and S. Barthelmy are with the Gooddard Space Flight Center, Greenbelt, MD 20771, USA. P. Mao and H. Miyasaka are with the California Technology of Institute, Pasadena, CA 91125, USA. J. Apple is with the Marshall Space Flight Center, Huntsville, AL 35812, USA.

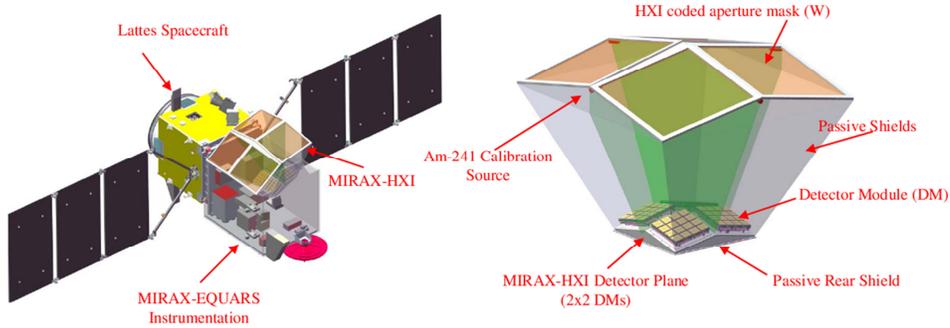

Fig. 1. Monitor e Imageador de RAios-X (*MIRAX*) is the first Brazil-INPE space astrophysics experiment on the Lattes satellite (*Left*), scheduled to launch in 2017. The Hard X-ray Imager (HXI, *Right*) on *MIRAX* is a wide-field hard X-ray survey telescope, and the current design consists of 2 × 2 identical coded-aperture telescopes mounted on top of the Equatorial Atmosphere Research Satellite (EQUARS) instruments.

flight in section V. Analysis of the imaging data is in progress and will be reported in a future publication.

## II. PROTOEXIST1 & 2 (P1 & P2)

*P1 & P2* are our on-going balloon-borne wide-field hard X-ray telescope experiments to advance the technology for CZT detector and coded-aperture imaging for next generation hard X-ray survey telescopes. *P1* is the first generation of the program, employing 16 cm × 16 cm of CZT detectors with 2.5 mm pixels [5,6]. *P2* is the second generation of the program, allowing the same active area of the CZTs detector but with 0.6 mm pixels. Table I summarizes the key telescope and detector parameters for both *P1* and *P2* during the 2012 flight.

TABLE I. KEY PARAMETERS FOR *PROTOEXIST1* (*P1*) & 2 (*P2*) DURING '12 FLIGHT

| Parameters | *P1* | *P2* |
|---|---|---|
| Sensitivity (5σ)* | ~140 | ~140 mCrab/hr |
| Energy Range | 30 – 300 | 5 – 200 keV |
| Energy Resolution | 3 – 5 | 2 – 3 keV |
| FoV (50% coding) | 20° × 20° | 20° × 20° |
| Angular Resolution | 20′ | 5′ |
| CZT Detector | 64 Crystals | 56 Crystals |
| Side Dimension | 1.93 cm | 1.98 cm |
| Active Area | 238 cm² | 220 cm² |
| Pixel Size | 2.5 mm | 0.6 mm |
| Thickness | 5 mm | 5 mm |
| Tungsten Mask | 12 × 0.3 mm | 4 × 0.1 mm |
| Pixel Pitch, Grid Width | 4.7, 0.5 mm | 1.1, 0.1 mm |
| Pattern | URA | Random |
| Mask-Det. Sep | 90 cm | 90 cm |
| Rear & Side Shields | Graded Pb/Sn/Cu for 5 sides each | |
| $^{241}$Am Cal. Source | 220 nCi each, ~36 cm above the det. | |

\* Assuming no atmospheric absorption.

Fig. 2 shows a picture of the *ProtoEXIST* payload, which was taken during a magnetometer vs. gondola Global Positioning System (GPS) calibration a few days before the flight. Both *P1* and *P2* detector planes sit inside of the pressure vessel (PV) along with the flight computer and other electronics. The *P1* and *P2* mask towers are mounted on the top panel of the PV. The daytime start camera, mounted at the end of the long baffle on the left is a primary instrument of aspect and pointing systems along with differential GPS, magnetometer, gyros, etc.

The main improvement from *P1* to *P2* is the four times finer spatial resolution of CZT detectors and the mask. As a result, *P2* can localize hard X-ray sources within 40″ accuracy or better at 90% confidence (1′ for *MIRAX*-HXI). This means that typical longer wavelength telescopes with a few arcmin scale FoV can follow up on a newly discovered source without an additional positional refinement to ensure that the FoV includes the source. Another major improvement is a lower energy threshold, which is ~ 5 keV for *P2* (and ~ 30 keV for *P1*). We give a more detailed description of the detector plane in the next section. Here we summarize other aspects of the telescopes.

In order to take advantage of the fine spatial resolution of the *P2* detectors, we also refined the mask pattern with 1.1 mm pixel pitch and 0.1 mm support grid. Fig. 3 compares the *P1* and *P2* masks. The masks are made of the stacked photo-chemically etched Tungsten sheets. The *P1* mask consists of 12 layers of 0.3 mm thick Tungsten sheets with 4.7 mm pixel pitch. The *P2* mask consists of 4 layers of 0.1 mm thick Tungsten sheets with 1.1 mm pitch. The precision limit in photo-chemical etching is driven by the thickness of the material (~110% of thickness), so precision etching of 0.1 mm grid requires 0.1 mm thick (or thinner) sheets. In *P2*, each layer consists of two sheets (20 cm × 40 cm piece each) placed side by side due to the size limitation of the 0.1 mm thick sheets by the manufacturer. Each sheet covers a half of the mask pattern. In order to maintain the precise pixel pitch across the sheets, the division of the pattern alternates its orientation (*x* vs. *y* direction for the 40 cm dimension) from layer to layer. We stacked four layers to provide sufficient modulation up to 200 keV X-rays (>90% attenuation of opaque pixels), and the four layers are held and supported by an X-ray transparent Mylar sheet (~100 μm) and an Al frame with central cross bars (6 mm wide).

A pattern of uniformly redundant array (URA) is used for the *P1* mask in order to minimize coding noise [8]. In *P2*, however, the relatively large gaps between close-tiled 2×2 CZT crystals (see section III) degrade the potential advantage of URA patterns. Since the number of pixels in the *P2* mask is relatively large (~64K) and the coding noise of random mask patterns is inversely proportional to the square root of the number of pixels, we employed a random pattern for the *P2* mask. The dynamical range, the maximum signal ratio of the brightest to faintest sources that can be detected simultaneously, of a random mask pattern with about 60K pixels is about 50 for detections at $5\sigma$ or greater (i.e. the maximum Signal-to-Noise Ratio (SNR) is about 250), whereas it would have been only about 13 for a random mask pattern of the *P1* pixel size.

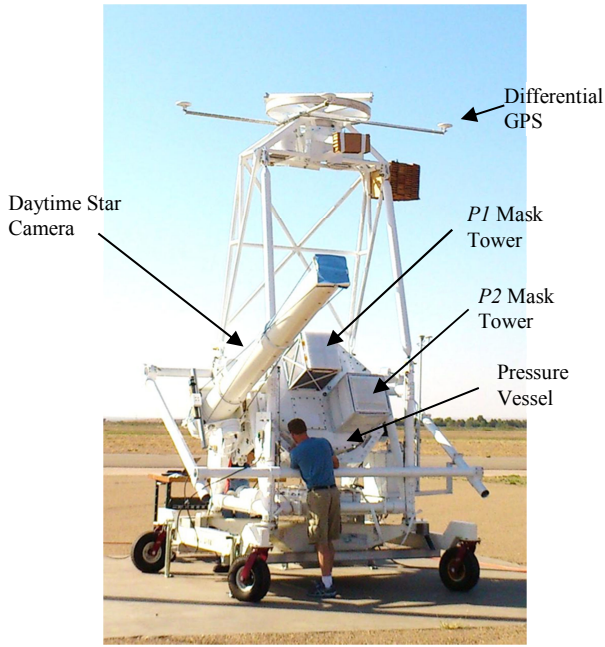

Fig. 2. *ProtoEXIST* Payload. Both *P1* and *P2* detector planes sit inside of the pressure vessel (PV) along with the flight computer and other electronics.

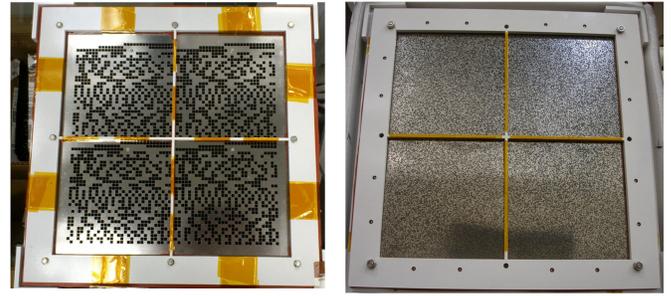

Fig. 3. The *P1* mask (*left*) consists of 12 layers of 0.3 mm thick Tungsten sheets etched with a URA pattern of 4.7 mm pixels. The *P2* mask (*right*) consists of 4 layers of 0.1 mm thick Tungsten sheets with a random mask pattern of 1.1 mm pixels.

We also redesigned the side and rear passive shield of both *P1* and *P2* for the 2012 flight. Following the current baseline configuration of *MIRAX*-HXI, we use passive shields for both side and rear shields instead of the CsI active rear shield used in the 2009 flight [6]. The passive shields are made of stacked layers of Pb/Sn/Cu sheets, which cascade down incoming background X-rays to lower energies below the threshold. Note at the float altitude of ~36–40 km, X-rays below 20 keV from cosmic sources are almost completely absorbed in the atmosphere, so events triggered with energies below 20 keV are atmospheric or electronics background. The side shields of each telescope are divided into two sections: one inside the pressure vessel (PV) and one outside of the PV, which reaches all the way up to the mask.

For onboard calibration of the spectral gain and dead time measurement during the flight, we mounted a 220 nCi $^{241}$Am radioactive source for each telescope at ~36 cm above the detector plane in the side shields.

## III. *P2* CZT Imaging Detector

For efficient packaging of a large active area of fine pixel CZTs, we modularized the design and assembly architecture of the *P2* detector plane as we did for *P1* [6]. In summary, the *P2* CZT detector consists of two vertically-stackable electronics boards, where 64 CZT crystals and matching 64 Application Specific Integrated Circuits (ASICs), each connected with 87 wirebonds to their ASIC Carrier Board (ACB), can be mounted (Fig. 4).

The basic modular block is a Detector Crystal Unit (DCU), an assembly of a 1.98 cm × 1.98 cm CZT crystal directly flip-chip bonded on an ASIC mounted on an ACB. We employ Redlen CZTs, which are shown to exhibit high quality at low cost. The crystal size is slightly larger than their standard unit, which allows a guard ring on the anode side for suppressing high edge leakage currents and potential degradation of the edge pixels. Their original contacts on both cathode and anode sides (8 × 8 pixels) put on by Redlen Technologies were stripped off and new metallization contacts with 32 × 32 anode pixels and new continuous cathode were sputtered on at Creative Electron INC (CEI) for hybridization.

For signal processing of the 1024 pixels on each DCU, we use Nu-ASIC, an ASIC developed for *NuSTAR* [9]. The Nu-ASIC, with a long heritage of development (including RadNET ASICs used in *P1*), comes with matching 1024 channels, whose input pad pattern is suitable for flip-chip bonding with a CZT crystal with 32 × 32 pixels at 600 μm pitch. A flexible readout scheme allows multi-pixel pulse profile retrieval for each event, which can be used for reconstruction of multi-trigger events such as Compton scattering and depth sensing using negatively charged neighboring pixels [10]. The Nu-ASIC is also equipped with a 12 bit analog-to-digital converter (ADC), and the internal digitization makes it less susceptible to electronic noise.

In a DCU assembly, we first bonded at SAMTEC gold studs on the 1024 input pads of a Nu-ASIC for later flip-chip bonding with a CZT crystal. Then we mount and wirebond the gold-studded Nu-ASIC on an ACB, which is already assembled with components and elastomer connector housing. Finally, we bond the package with a properly metalized CZT crystal using conductive epoxy at CEI. For speedy assembly of large number of DCUs, we bond and then cure 4 DCUs at once in a thermal chamber. Given the fine pixel pitch, we maintain the alignment of each assembly stage with a tolerance level of 50 μm or less, using alignment pins and precise visual inspection under microscope during the bonding procedure.

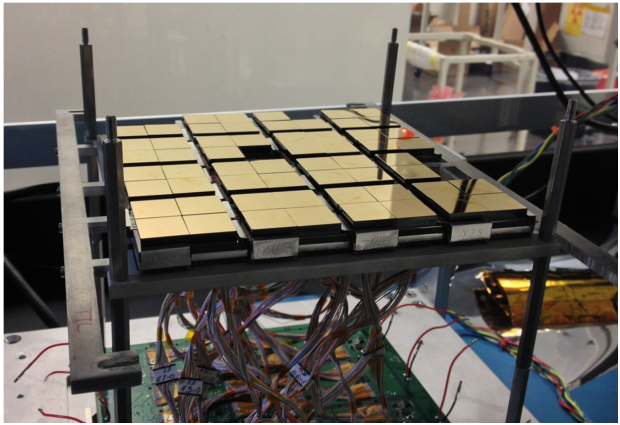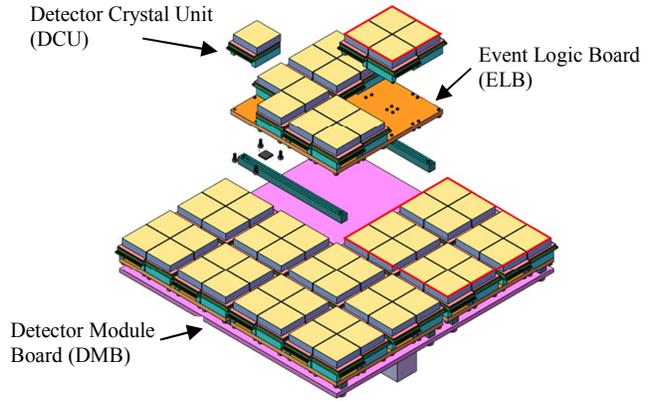

Fig. 4. The assembled *P2* detector plane (*left*) and its packaging architecture (*right*). The *P2* detector plane consists of two stackable boards (ELB and DMB), where CZT crystals and Nu-ASICs are mounted.

The next stage of the assembly module is based on an Event Logic Board (ELB), which contains 16 Complex Programmable Logic Devices (CPLDs) to handle and relay signals and commands between 16 DCUs and the Detector Module Board (DMB). The electric and mechanical connection between an ACB and the matching ELB is done by two elastomer connectors. For the mechanical alignment of the whole detector plane, 4 ELBs are mounted on a rigid Al frame through alignment pins. Some of the alignment pins are common to both an ELB and a DCU, and the rest are unique to each DCU. This enables pixel alignment of a CZT to another within ~100 μm.

As shown in Fig. 3, the CZT crystals in a 2×2 DCU set are closely-tiled with only 300 μm gaps, which, combined with the spacing needed for a guard ring on the anode side of each CZT crystal, results in a one pixel wide gap between adjacent detectors. Due to the real estate required for wirebonds of the Nu-ASICs, each set of 2×2 DCUs are mounted with a larger gap (12 pixels).

The full detector plane is assembled on the DMB (Fig. 4), where all the data is collected. The electric connection between the ELBs and the DMB is done through high density ribbon cables. The DMB contains a Xilinx Vertex-4 commercial grade Field Programmable Gate Array (FPGA) to operate and process all 64 DCUs. In addition, two EMCO High Voltage (HV) Power Supply (PS) units are controlled through the FPGA in the DMB. Each HV PS can bias 32 DCUs. The –600 V bias voltage is applied through Al tape that extends and covers a large portion of the CZT cathode surfaces to reduce the field distortion.

We read out the data from the DMB to a flight computer through a Low Voltage Differential Signal (LVDS) cable. A flight computer handled both the *P1* and *P2* data stream and passed the commands from the ground during the flight. All the raw data were saved on an onboard solid state disk, and a part of compressed data along with various housekeeping information were telemetered to the ground during the flight for monitoring of the detector and telescope systems.

Many design choices such as selection of the FPGA type and the link between the flight computer and the DMB were made under the consideration for space application (*MIRAX*). The overall yield of DCU assembly is > 80% with >90% of each stage (e.g. ASIC yield, bonding success, etc.), which is very promising for future programs requiring a large number of CZT detectors. Note that as seen in Sec V.C (Fig. 7), some detectors (~7%) show a large section of missing pixels. The severity of this issue appears to vary with the detector configuration (e.g. side shields), so we do not count them as failure in this yield calculation.

Due to unexpected non-technical delays in the program, we were only able to assemble 61 flight-ready DCUs and mounted 58 DCUs out of 64 slots before the flight as shown in Fig. 3. Three DCUs also show a deviation in the mechanical alignment larger than 150 μm, which kept them from the final assembly (the nominal gap between CZT crystals is only 300 μm). This is in part due to uneven wall polishing of these CZT crystals, which can be better controlled in future.

## IV. PRE-FLIGHT CALIBRATION

After integration of the science payload and gondola, we conducted a boresight calibration to measure the pointing offset between the two X-ray telescopes and the daytime star camera. The daytime star camera can reach $V_{mag}$ ~ 10 with a FoV of 2º×3º.

We placed a 10 mCi $^{241}$Am source at 15.8 m away from the X-ray telescopes with an optical target separated from the $^{241}$Am source by the separation of the star camera from the midpoint of *P1* and *P2*. Fig. 5 shows the *P2* detector plane image and the reconstructed sky image of both the *P1* and *P2* X-ray data from the boresight calibration. The detector plane image is from single pixel trigger events in the 50 to 65 keV band. The image shows the shadow of the central cross bars of the mask frame, and the data are missing for 6 empty slots and 2 malfunctioning detectors.

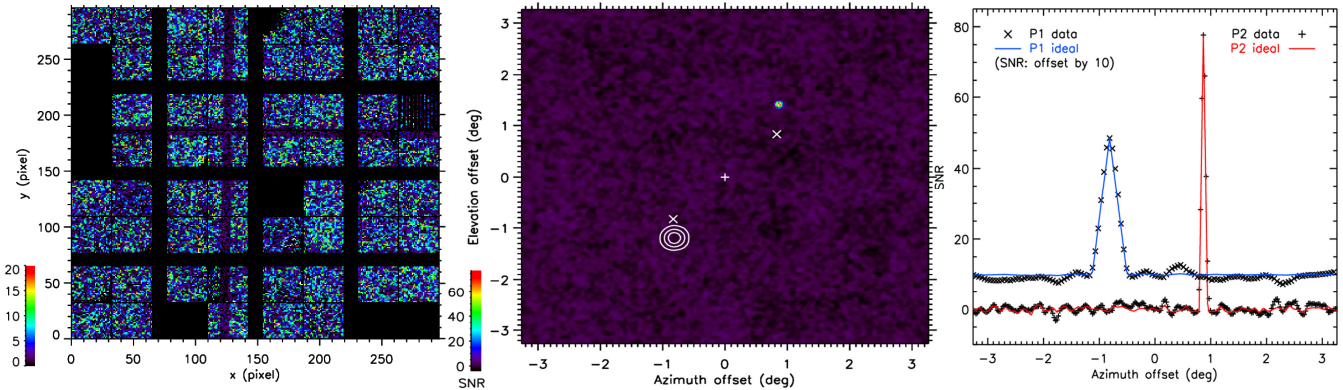

Fig. 5. An event distribution in the *P2* detector (left), the reconstructed sky images (middle) of *P1* (contour) and *P2* (color map) from the boresight calibration data using an external [241]Am source, and a 1-D cross sectional view of the sky images (right) in comparison with ideal point spread functions (PSFs; the peak values are normalized to the data).

In the reconstructed sky image (the middle panel), the color map is from the *P2* data, and the contour (SNR=10, 20, & 30) is from the *P1* data. For imaging, we corrected for the rotational offset of the mask pattern relative to the detector pixel axes (0.6° for *P1*; 0.74° for *P2*), which is visible in the cross bar shadow in the detector image. The observed SNR of the source is ~77 for *P2* and ~42 for *P1*, and both are limited by the coding noise of the system and non-uniformities of detectors. A drastic improvement (4×) of the spatial resolution of *P2* relative to *P1* is also clear in the image. The "+" and "x" symbols indicate the pointing direction of the star camera before and after accounting for the separation of the external [241]Am source and the optical target next to the source respectively (note two "x"s for two telescopes). Once the parallax due to the finite distance of the external source is accounted for, the boresight offset of the X-ray telescopes relative to the star camera is found to be (Δaz, Δel) = (+4′, -25′) for *P1*, and (+5′, +39′) for *P2*, which is roughly what we expected from the uncertainty of various stages of the alignments in the integration.

The right panel in Fig. 5 shows a 1-D cross sectional view of the sky images in comparison with ideal point spread functions (PSFs; the peak values are normalized to the data). Note that even the ideal PSFs exhibit noise fluctuation (coding noise) due to the random nature of the mask pattern in the case of *P2*. The ideal *P1* PSF of the boresight source also shows small coding noise originating from an incomplete sampling of a cycle of the URA mask pattern due to image magnification caused by the finite source distance. The defects in our detectors lead to a degradation in source positioning mostly through signal reduction and noise increase rather than the PSF blurring since the average pixel distortion scale is still less than a half of the mask pixel size [7].

## V. Detector Performance in Flight

The successful reconstruction of the external calibration source has demonstrated that both the *P1* and *P2* telescopes operate properly. After a few launch attempts cancelled due to high surface winds, on 2012, Oct 10, we had a successful launch of the payload with a 29 M cubic feet balloon. After 2 hours of climb-up, the payload reached the float altitude of 39 km, where the flight lasted about 8 hours before termination.

Throughout the flight both the *P1* and *P2* detectors performed as well as on the ground. Only a relatively small number of hot pixels developed and were subsequently disabled during the flight. We observed relatively bright X-ray targets such as Sco X-1, GRS 1915+105, Swift 1745.1-2624, and Cyg X-1, spending ~30 min to 1 hr each. The PV kept the temperature of the detector planes in between ~10 and 20 ºC during the flight. The thresholds of the *P1* and *P2* detectors were set at ~30 and 18 keV respectively. For *P2*, near the end of the flight after the targeted observations, we experimented with detector settings such as threshold and readout mode. The *P2* detector operated nominally at a threshold of ~6 keV. In-depth analyses of the flight data are underway. Here we present the initial results to evaluate the performance of the *P1* and *P2* telescopes.

### A. Event Rate

The left panel in Fig. 6 shows the event rate history of the *P2* detector from the prelaunch warm-up to the end of the flight. The time (*x*-axis) is calculated from when the instrument was initially turned on. The payload was launched at 2 hr (~8:35 am local time). The black line indicates the overall event rate in the 20 to 200 keV band. The red and orange lines show single pixel and multi-pixel trigger events in the same energy band respectively. The energy value of each event is calibrated using the onboard [241]Am calibration source (see also section IV-B). The blue line shows [241]Am calibration source event rates, scaled (8×) for easy reading. The calibration source event rates are selected from single pixel trigger events in the 55 to 65 keV band after background subtraction using the same in the 75 to 85 keV band.

The event rates rose as the payload climbed up after a brief drop in the beginning. As the payload climbed above the Pfotzer maximum (~20 to 25 km), the event rates started dropping after peaking at 400 cps in the 20 to 200 keV band. By the time the payload reached float altitude (39 km), the event rates settled at around 280 cps, which remained more or less steady through the rest of the flight. In order to extract the true X-ray incident rate, many corrections need to be applied. For instance, the variation of the onboard calibration source event rates (blue) relative to the overall event rate (black) basically illustrates the rate-dependent dead time variation.

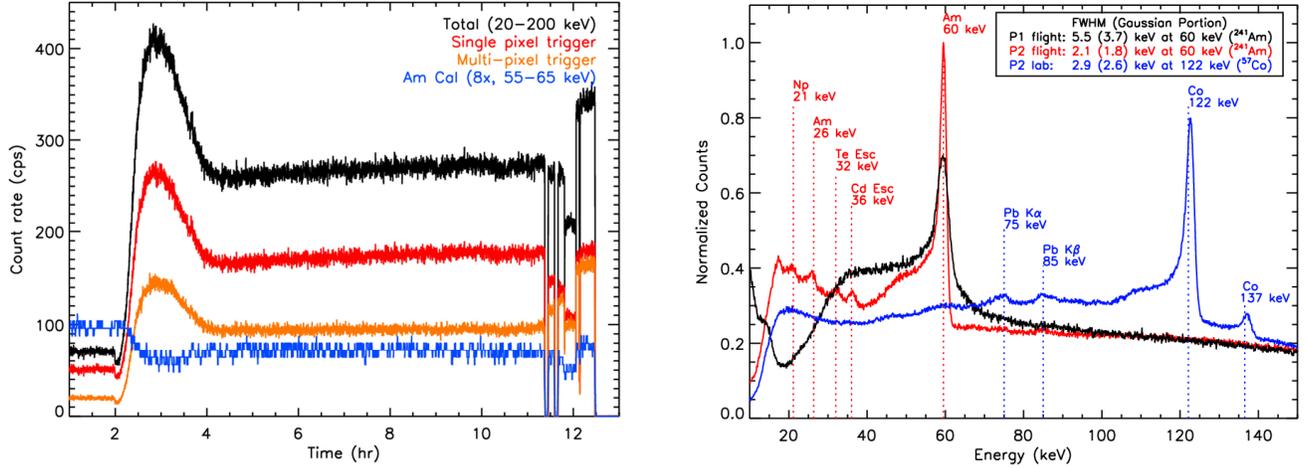

Fig. 6. The various types of event rate history (*left*) and the combined spectrum (*right*) of the *P2* detector during the flight. The *P2* flight spectrum (red) is compared with the *P1* flight (black) and *P2* ground (blue) spectra.

The sudden variation of the event rates near the end of the flights is due to the experimental changes of detector settings – readout modes and thresholds. For a given trigger pixel in each event, we normally read out the immediate surrounding pixels as well as the trigger pixel [5,10]. For instance, for a non-edge pixel trigger, we read out 9 pixels in total, which is called 3×3 mode. This multi-pixel readout enables recovery of charge split over adjacent pixels, whose energy values can be recorded even below the threshold in some cases. At the same time, it also allows depth measurement of X-ray events, which can be used for depth-dependent background subtraction. During the last hour of the flight, we changed the readout mode from 3×3 to 5×5 and to 1×1 at the very end. The 1×1 mode reads out only triggered pixels. For each pixel we read out 16 sampling points in the pulse profile [5,6,10] and the dead time increases as the number of readout pixels increases. Comparison of the event rate with 1×1 and 3×3 readout mode indicates the dead time fraction in the normal 3×3 mode is about 20% or more. This is roughly consistent with the ground-to-flight variation (~25% drop) of the event rate of the calibration source (blue). The ratio of single (red) to multi-pixel (orange) trigger event rates appears to depend on the readout mode.

### B. Energy Spectra

The right panel in Fig. 6 shows the combined spectra of the single pixel trigger events during the flight. The *P2* flight spectrum (red) is compared with the *P1* flight (black) and *P2* ground (blue) spectra using a $^{57}$Co source. Even with the 8 hr integration during the flight, the large number of pixels (~57k) and a relatively weak onboard calibration source (220 nCi) result in an order of 10 counts per pixel in the 55 to 65 keV band, which is not sufficient for precise pixel-by-pixel calibration. Therefore, the *P2* spectra are corrected for the pixel-to-pixel variation of spectral gain, based on a long post-flight ground measurement of X-rays from a 10 mCi $^{241}$Am source. All the major low energy lines including escape peaks are clearly visible.

For the *P1* spectra, capacitor ID-based correction is also applied in addition to pixel-to-pixel correction [5,6,10], also based on a separate post-flight ground measurement of X-rays from the 10 mCi $^{241}$Am source. Note that in both *P1* and *P2*, for a given pixel, X-rays with the same energy can produce different signals in the ASICs, depending on the sampling order of pulse profiles by 16 capacitors [10]. However, unlike RadNET ASICs used in *P1*, in Nu-ASICs of *P2*, the capacitance values of sampling capacitors is more uniformly controlled, so the capacitor ID-based correction is not necessary for *P2* especially under the normal operational mode.

Two spectral resolution (FWHM) values are shown for each spectrum: one for the full range and the other for the Gaussian portion of the peak excluding the contribution from the exponential low energy tail [10]. For instance, the *P2* flight spectra shows 2.1 keV FWHM for the full range and 1.8 keV FWHM for the Gaussian portion. This means the high energy side of the FWHM around the peak is 0.9 keV and the low energy side is 1.2 keV. The larger low energy tail and shoulder in ~45–55 keV are mainly due to charge split events (but triggering only one pixel) rather than depth-dependent variation since 60 keV X-rays mostly interact within 1 mm from the cathode. A more sophisticated analysis algorithm of the 3×3 mode data taking into account charges in the non-triggered neighbor pixels can properly reassign the energy of the events in the shoulder [e.g. 10], but simply lowering the threshold can also reduce the low energy shoulder by registering some single-pixel trigger events properly as multi-pixel events.

In the case of *P1*, the FWHM resolution is 5.5 keV at 60 keV and the Gaussian portion is about 3.7 keV, therefore the *P2* detectors show a factor of two improvement in resolution. This improvement in the spectral resolution is also reflected in the lower threshold setting capability of *P2* (~6 keV vs. 30 keV for *P1*).

Fig. 6 also shows the ground *P2* spectrum taken from a $^{57}$Co source, indicating 2.9 keV FWHM at 122 keV. The Pb fluorescence lines at 75 and 85 keV in the $^{57}$Co spectrum are from the source holder made of lead. The background in the $^{57}$Co spectrum is relatively high for a lab measurement due to the relatively weak activity (< 100 µCi) of our $^{57}$Co source. Since the $^{57}$Co spectrum is also calibrated based on the 60 keV X-ray line from the $^{241}$Am source and a linear extrapolation to higher energy X-rays, a slight non-linearity in the gain is visible in the spectrum at the level of ~0.5 keV at ~120 – 130 keV.

TABLE II. RUNDOWN OF ELECTRONICS NOISE & ENERGY RESOLUTION

| Stage | *P1* (keV, FWHM) | *P2* (keV, FWHM) |
|---|---|---|
| Electronics Noise | 2.1 * | 1.1 |
| +Hybridization | 2.5 | 1.6 |
| +Leakage Current (–600V) | 3.0 | 1.7 |
| Resolution at 60 keV (Gaussian portion only, with low energy tail) | | |
| During flight '12 | 3.7, 5.5 | 1.8, 2.1 |
| References | [5,6] | [7] |

The non-flight measurements are based on a few selected detectors. *This already includes a part of the capacitance noise from the traces in the interposer board [5].

Table II shows a rundown of electronics noise and energy resolution in the *P1* and *P2* detectors along each stage of the assembly to illustrate noise contribution of each component. For instance, the leakage current contributes about 0.6 keV in the noise of the *P2* detectors, based on the noise increase from 1.6 keV to 1.7 keV by applying a bias voltage of –600V and the assumption that all the contributions are independent to each other (i.e. quadratic summation). In *P1*, the leakage current contributes about 1.7 keV. Note that each *P2* CZT crystal has 1024 pixels as opposed to 64 pixels in *P1* under a similar footprint, so that with the same crystal, the Nu-ASICs in *P2* would experience a 16 times lower leakage current than the RadNET ASICs in *P1*. Therefore, the noise contribution of the leakage current is less significant in *P2* than in *P1*. Similarly the noise increase due to hybridization indicates the contribution from the input capacitance noise of crystal bonding, although in *P1*, there is an additional input capacitance noise due to the connecting traces in the interposer board.

Note that the temperature variations in the system between the ground (usually ~23–28° C) and the flight measurements (~8–18°C) drive the resolution change in the *P2* spectra. For instance, the FWHM resolution on the ground is about 2.2 keV at 60 keV (not shown in the figure) instead of 2.1 keV during the flight, so we expect a slightly better resolution at 120 keV during the flight than what is shown in Fig. 6. In the case of the *P1* detector, the gain changes with the temperature (~0.11 keV/°C) while the resolution show little variation since the non-thermal noise components dominate. In the *P2* detector, the gain does not vary noticeably thanks to the active thermal correction feature in Nu-ASIC, but the resolution variation is now noticeable since the thermal noise component is comparable to other source of the noise.

The *NuSTAR* detectors show ~1 keV FWHM at 60 keV with ~2 keV threshold [9]. Among many subtle (but possibly critical) differences between the *NuSTAR* and our detectors, selection criteria of higher detector quality in general, which *NuSTAR* can afford (*NuSTAR* needs only 8 DCU equivalents vs. 64 DCUs in *P2* or 256 in *MIRAX*), likely contribute to a better overall performance of the NuSTAR detectors. But we suspect that the main difference in the spectral performance is originated from the different operational modes of Nu-ASIC. While we ran Nu-ASICs in the normal mode, *NuSTAR* runs Nu-ASICs in the charge-pump (CP) mode. The latter corrects more aggressively for leakage currents, etc, thus improving the resolution and threshold. To meet the tight flight schedule, we only had time to implement the normal mode. The charge pump mode requires a non-trivial FPGA code change. We plan to implement the CP mode in our system in the future, so there is room for further improvement.

Before the flight, we have identified and disabled 609 hot pixels in 56 DCUs (1% of 57K pixels; also note 2 DCUs malfunctioned). The flight data indicates the additional 773 pixels were mildly hotter than the rest. When we exclude events from these pixels (~1.4% of the total), we can lower the low energy background by 30% (not shown). In coded-aperture imaging, 2.4% loss of pixels means about 1.2% reduction in SNR. A simple depth sensing cut using the neighbor pixel charges [5,6,10] also reduces the background by ~25 % at low energies (<40 keV) and ~50% at high energies (>60 keV).

### C. Event Distribution in the Detector

Fig. 7 shows the event distribution of the *P2* detector during the flight. The distribution is generated from the single pixel trigger events in the 20 to 200 keV band, and the image is drawn with two types of gaps between the CZT crystals with no correction for small pixel distortion. There are 6 DCUs missing and 2 DCUs malfunctioning, one of which generated a strip pattern in the event distribution (the other did not produce any events).

The event distribution immediately reveals a few interesting features. First, there is an enhancement of events in the edge pixels of each CZT crystal. This feature is more clearly visible when the data are folded onto the coordinates of a DCU, as shown in the right panel of Fig. 7. The edge event excess is due to the additional X-ray entrance surface for edge pixels through the side walls of each CZT crystal and the quasi-isotropic instrumental background such as Compton-scattered X-rays from surrounding material. In addition, a thick Al housing of a rectangular ring shape for elastomer connector right underneath each ACB is suspected to be a source of the event enhancement in the next-to-edge pixels relative to the inner pixels. Incomplete collection of multiple pixel trigger events on the edge pixels can also artificially enhance edge pixel events registering as fewer pixel trigger events. For instance, many single-pixel trigger events in the edge pixels are in fact double or triple trigger pixel events with missing neighbors.

Second, some DCUs lack events in pixels near a corner of the crystals. This might be due to degradation in the bonding of pixels near the corner or an extreme version of small scale pixel distortion, which is likely originated from the electric field distortion near the edge of the CZT crystals. In the former, if a large number of the pixels lose the electric connection with the ASIC, the nearest adjacent pixels may exhibit a significant degradation in the energy resolution or become substantially noisier due to re-routed additional leakage current. In the latter, we expect that the effect would be dominant for the outer edge of the DCUs on the outer-most edge of the detector plane, but the data show that it is not the case except for only one DCU. In fact, the exact cause of the small pixel distortion (e.g. whether it is non-uniformity of the metallization contact or intrinsic to the crystal properties) is still under investigation. For instance, pixel distortion is often seen more severe along the edges of the crystals rather than near the corners where the electric field distortion would be most severe. In addition, apparent solutions of the edge pixel distortion such as extending cathode planes or grounded side shields around each crystal [7] also require a more rigorous physical explanation and may not work for all the cases. For instance, A. Bolotinikov informed us that extending cathode does not improve the pixel distortion of their detectors [11].

Surface properties of crystals, the external environment (e.g. humidity), or long term stability of detectors (e.g. bonding) have been also suggested to play some roles. In the case of surface properties, polishing quality of crystal walls along with crystals' internal structure may explain seemingly diverse patterns among the detectors. A more definite proof can come from re-polishing (and re-assembling) one of the detectors with severe pixel distortion along the edges.

The external environment does not appear to be an issue since operations in the lab bench (humidity level ~30-50%) or in the $N_2$-filled PV did not make a noticeable difference in the pixel distortion. Long term stability does not explain the result: pixel distortion is present in the very first data set taken right after the assembly, and since the full assembly of the detector about a year ago, none but one detector show noticeable change in the performance so far. Only one detector lost response to radiation in ¾ of the pixels, but the exact cause is still under investigation (e.g. bonding degradation or damage during the landing of payload). See [7] for more about the pixel distortion and performance of these detectors; we are continuing to work on these measurements and will report new results when they are available.

Third, in several DCUs, the event distribution seems to show a pattern (streaks, patches), which might originate from the crystal structure. During the assembly procedure, we have acquired near-infrared (nIR) images (900-1100 nm) of all the CZT crystals we have used. These images did not reveal many structures except for a couple of crystals. This may be because the resolution of the nIR image is not sufficient to reveal any structural anomalies in these crystals, or because the peculiar pattern of the event distribution is not related to the crystal structure (e.g. a secondary effect of the pixel distortion). Further investigation is needed.

## VI. SUMMARY AND FUTURE WORK

We have assembled a large area (220 cm$^2$) of fine pixel (0.6 mm) CZT imaging detector (*P2*) for a balloon-borne wide-field hard X-ray telescope and next generation wide-field hard X-ray telescopes such as *MIRAX*-HXI. Compared to the previous *P1* detector, we have achieved a factor of two improvement in spectral resolution (2.1 keV at 60 keV) and a factor of five reduction in threshold (6 keV) in *P2*. The *P2* telescope also introduces high resolution hard X-ray coded-aperture imaging with a factor of four improvement in angular resolution (5′).

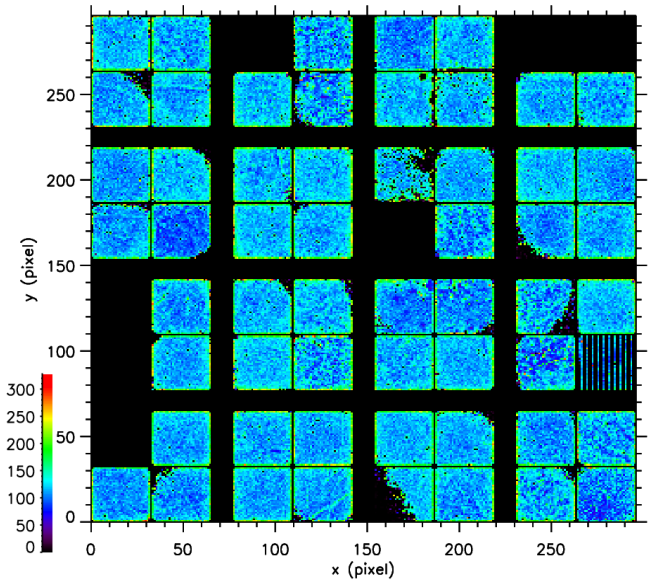
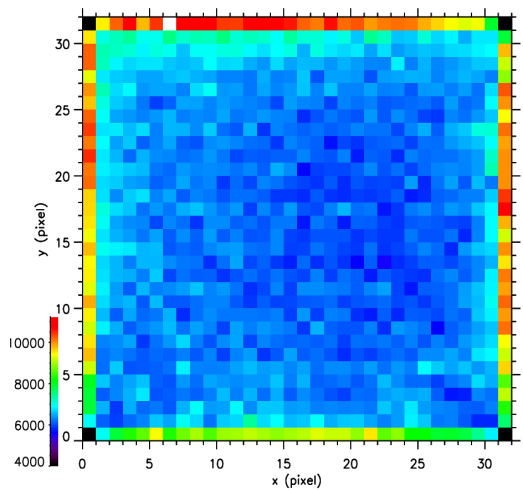

Fig. 7. The distribution of the single pixel trigger events in the *P2* detector during the flight (*left*) and the same folded onto the coordinates of a DCU (*right*).

We presented the first results of the high altitude balloon flight data, which illustrates the successful operation of the *P2* detector in the near space environment. The detector and telescope systems including the aspect and pointing system worked well during the flight. An in-depth analysis of the pointing and aspect data as well as the *P1* and *P2* data to generate X-ray images of the observed sources is underway.

ACKNOWLEDGMENT

We thank the Columbia Science Balloon Facility (CSBF) team for their excellent work and support of the balloon flight from scheduling the flight to the recovery of the payload. We also thank W. Cleveland, K. Dietz and D. Huie for their support for the flight. We thank M. Burke, N. Gehrels, W. R. Cook and F. Harrison for their support in the development and assembly of the *P2* detector.

REFERENCES

[1] Jonanthan E. Grindlay, "GRBs probes: increasing both the high-z and short GRB sample", 2012, Memorie della Societa Astronomica Italiana Supplement, v.21, pp. 147-154
[2] B. Abbott, et al.,"Detector description and performance of the first coincidence observations between LIGO and GEO," 2004, Nucl. Instrum. Methods A, 517, pp. 154–179
[3] C. Winkler, et al., "The *INTEGRAL* mission," 2003, A&A, 411L, pp. 1-6
[4] N. Gehrels, et al., "The *Swift* gamma-ray burst mission," 2004 ApJ, 611, pp. 1005-1020
[5] J. Hong, et al., "Building large area CZT imaging detectors for a wide-field hard X-ray telescope - *ProtoEXIST1*", 2009 Nucl. Instrum. Methods A, 605, pp. 364-373
[6] J. Hong, et al., "Flight performance of an advanced CZT imaging detector in a balloon-borne wide-field hard X-ray telescope—*ProtoEXIST1*," 2011 Nucl. Instrum. Methods A, 654, pp. 361-372
[7] Branden Allen, et al., "Development of the *ProtoEXIST2* advanced CZT detector plane," 2011 IEEE (NSS/MIC), pp. 4470 - 4480
[8] A. Busboom, H. Elders-Boll, & H. D. Schotten, 1998 Experimental Astronomy 8, 97
[9] Fiona A. Harrison, et al., "The Nuclear Spectroscopic Telescope Array (*NuSTAR*) High-energy X-ray Mission," 2013, ApJ, 770, pp. 103-121
[10] J. Hong, et al., "CZT Imaging detectors for *ProtoEXIST*," 2006, SPIE 6319, pp. 63190S-1-11
[11] A. Bolotinikov, 2012, private communication